\documentclass{article}
\usepackage{spconf,amsmath,graphicx}
\usepackage{amsfonts,amsmath}
\usepackage{amssymb,latexsym}
\usepackage[iso-8859-7]{inputenc}
\usepackage{psfrag}

\title{Retinal-inspired filtering for dynamic image coding}
\name{$\text{Effrosyni Doutsi}^{1,2}$, $\text{Lionel Fillatre}^{1}$, $\text{Marc Antonini}^{1}$ and $\text{Julien Gaulmin}^{2}$ \thanks{Thanks to 4G-TECHNOLOGY and ANRT for funding.}}
\address{$^{1} \text{Univ. Nice Sophia Antipolis, CNRS, I3S, UMR 7271, 06900 Sophia Antipolis, France}$ \\ $^{2} \text{4G-TECHNOLOGY, 460 avenue de la Quiera 06370 Mouans Sartoux - France.}$}
\usepackage{amsthm}
\usepackage{appendix}
\usepackage{multirow}
\usepackage{bigstrut}
\usepackage{textcomp}
\usepackage{bbm}
\usepackage{graphicx}
\usepackage{caption}
\usepackage{subcaption}
\usepackage{makeidx}
\makeindex

\newcommand{\SUM}[2]{\displaystyle\sum_{#1}^{#2}}

\newcommand{\Gc}{G_{\sigma_{c}}}
\newcommand{\Gs}{G_{\sigma_{s}}}
\newcommand{\Rc}{R_{c}(t)}

\newcommand{\Rs}{R_{s}(t)}

\newcommand{\R}{{\mathbb{R}}}	
\newcommand{\Rn}{{\mathbb{R}}^{2}}	           

\newcommand{\DKxt}{\phi(x_{k}-x_{i}, t_{j})}			

\newcommand{\Txt}{T(t-t')}
\newcommand{\TExt}{(T\overset{t}{*}E_{\tau_{S}})(t-t')}
\newcommand{\Dfx}{f(x_{i})}	 														
\newcommand{\Dt}{\Delta t}								
\newcommand{\intt}{\displaystyle \int_{t'=0}^{t}}

\newcommand{\defk}[1]{k = 1,\ldots ,n }     									
\newcommand{\defj}[1]{j = 1,\ldots ,m }		 					   		    
\newcommand{\defi}[1]{i = 1,\ldots ,n }       								    

\newtheorem{prop}{Proposition}[section]

\begin{document}
\maketitle
\begin{abstract}
This paper introduces a novel non-Separable sPAtioteMporal filter (non-SPAM) which enables the spatiotemporal decomposition of a still-image. The construction of this filter is inspired by the model of the retina which is able to selectively transmit information to the brain. The non-SPAM filter mimics the retinal-way to extract necessary information for a dynamic encoding/decoding system. We applied the non-SPAM filter on a still image which is flashed for a long time. We prove that the non-SPAM filter decomposes the still image over a set of time-varying difference of Gaussians, which form a frame. We simulate the analysis and synthesis system based on this frame. This system results in a progressive reconstruction of the input image. Both the theoretical and numerical results show that the quality of the reconstruction improves while the time increases.
\end{abstract}
\section{Introduction}
\label{sec:intro}

During the last decades there has been a great progress in image and video compression standards which enables to deal with the current needs of coding and decoding High Definition (HD) and Ultra High Definition (UHD) signals \cite{Poynton03}. However, this progression rate  seems to be very low comparing to the increasing rate of the amount of data which needs to be transmitted or stored. For instance, H.265/HEVC \cite{Grois13} was released in 2013, ten years later than the previous standard H.264/AVC \cite{Gao13,Richardson11}. This lack of synchronization restricts their evolution which is currently based on parameterizations and/or improvements of the basic architectures instead of the proposition of groundbreaking approaches.

In this paper we study an alternative compression and decompression model, which is based on the behavior of the visual system. The way the luminance of light is captured, transformed and compressed by the inner part of the eye, the retina, seems to follow the basic principles of compression. The retinal function has been explicitly modeled by neuroscientists and the experimental results have shown that this should be a very efficient ``compression" model \cite{Wohrer09}. This is due to the fact that the retina is a layered structure of different kinds of cells. The amount of cells decreases while they are closer to the optic nerve. This structure succeeds in encoding the information in order to fit the limited capacity of the optic nerve. The signal which reaches the eyes is successfully transmitted to the brain despite this bottleneck \cite{Gollisch10, Masland01}.

Our goal is to study the retinal-inspired transformation from the signal processing point of view and to set the basis for our future bio-inspired dynamic codec. The first attempt in modeling this kind of filter was proposed in \cite{Masmoudi13}.  This is a separable spatiotemporal filter structured as a Difference of Gaussians (DoG) pyramid based on \cite{Burt83,Field94}. Each layer of this structure is delayed with an exponential temporal function. We improve the filter by explicitly taking into account the time in the design of our non-Separable sPAtioteMporal (non-SPAM) filter. 

In coding/decoding systems, the transformation of the signal is extremely important because it results in a more suitable representation in terms of the number of informative coefficients. There are different kinds of transformations like Discrete Fourier Transform (DFT) \cite{Mallat99}, Discrete Cosine Transform (DCT) \cite{Ahmet74} and the Discrete Wavelet Transform (DWT) \cite{Daubechies90, Antonini92, Vetterli95, Gupta15}, which are currently used in most of the lossy conventional compression standards (i.e., JPEG and JPEG2000). In addition, there are other kinds of filters which have been built in order to serve not only image compression but the general aspects of image and video processing. For instance, the Gaussian and Laplacian pyramids \cite{Burt83} are scaled spatial filters which allow a progressive transmission of a signal. Other kinds of pyramids are for instance the oriented pyramids using Gabor functions \cite{Watson87} or the orthogonal pyramids \cite{Adelson88}. Many of these approaches taking into account the time has been extended into spatio-temporal filters which have been applied especially in video surveillance and object tracking techniques \cite{Leduc97, Choi97}.  

In section  \ref{sec:filter} we aim to introduce the non-SPAM filter and explain its bio-inspired nature. We also applied this filter on a still image which is flashed for a long time. Then, we prove that the non-SPAM filter has a frame structure in section \ref{sec:frame}. In section \ref{sec:reconstruction}, we propose a progressive reconstruction of the input signal which is numerically illustrated in section \ref{sec:results}. In the last section, we conclude the paper with a discussion about the future work.

\section{non-SPAM Filter}
\label{sec:filter}

The aim of this section is to introduce the non-SPAM filter and to study its behavior. This filter is inspired by the mechanism of photo-receptors and horizontals cells which lie inside the retina (the innest part of the eye). These cells act as edge detectors and at the same time as motion detectors due to the way they connect to each other. These are the features that the non-SPAM filter tries to mimic having a spatial behavior which varies with respect to time. The non-separability of space and time enables the filter to detect temporal variations of luminance even in a uniform spatial region. This is not the case for a separable spatiotemporal filter. We study the non-SPAM filter for the analysis and synthesis of an image $f(x,t)$, where $ x\in \Rn$ and $t\in\mathbb{R^{+}}$, which is observed during a certain time interval. The spatiotemporal convolution of the non-SPAM and the input image results in the function $A(x,t)$ called the activation degree:
\begin{equation}\label{eq:SpaTempConv}
A(x,t) = K(x,t) \overset{x,t}{*} f(x,t)
\end{equation}
where $\overset{x,t}{*}$ is the convolution with respect to space and time.  We are going to define now the non-SPAM filter  in continuous time and space as following:
\begin{equation}
\begin{array}{l l l l}
K(x,t) &= C(x,t) - S(x,t), &
\label{eq:9}
\end{array}
\end{equation}
where $C(x,t)$ and $S(x,t)$ are the center and the surround spatiotemporal filters given by equations (\ref{eq:center}) and (\ref{eq:surround})  respectively:
\begin{equation}
\begin{array}{l l l l}
C(x,t) &= w_{c}\Gc (x) T(t), &
\end{array}
\label{eq:center}
\end{equation}
\begin{equation}
\begin{array}{l l l l }
S(x,t) &= w_{s}\Gs(x) \left( T \overset{t}{*}E_{\tau_{S}}\right)(t),
\label{eq:surround}
\end{array}
\end{equation}
where  $w_{c}$ and $w_{s}$ are constant parameters, $\Gc$ and $\Gs$ are spatial Gaussian filters standing for the center and surround areas respectively, and  $E_{\tau_{S}}$ is an exponential temporal filter. The center temporal filter $T(t)$ is given by:
\begin{equation}
T(t)  = E_{\tau_{G},n} \overset{t}{*} \left( \delta_{0} - w_{c}E_{\tau_{C}} \right) (t),
\label{eq:tempCenter}
\end{equation}
where the gamma temporal filter $E_{\tau_{G},n}(t)$ is defined by
\begin{equation}
E_{\tau,n}(t) =  \dfrac{t^{n} \exp\left( -t/ \tau\right)}{\tau^{n+1}},
\label{eq:gamma}
\end{equation}
where $n\in \mathbb{N}$ and $\tau$ is a constant parameter ($E_{\tau,n}(t)=0$ for $t<0$), $E_{\tau_{C}}$ is an exponential temporal filter, $\delta_{0}$ is the dirac function and $\overset{t}{*}$ stands for the temporal convolution. In case that $n=0$, the gamma filter turns to an exponential temporal filter. The convolution of the temporal filter $T(t)$ with the exponential filter $E_{\tau_{S}}$ is related to the delay in the appearance of the surround temporal filter with respect to the center one.

\begin{figure}[!ht]
\centering
\includegraphics[scale=0.40]{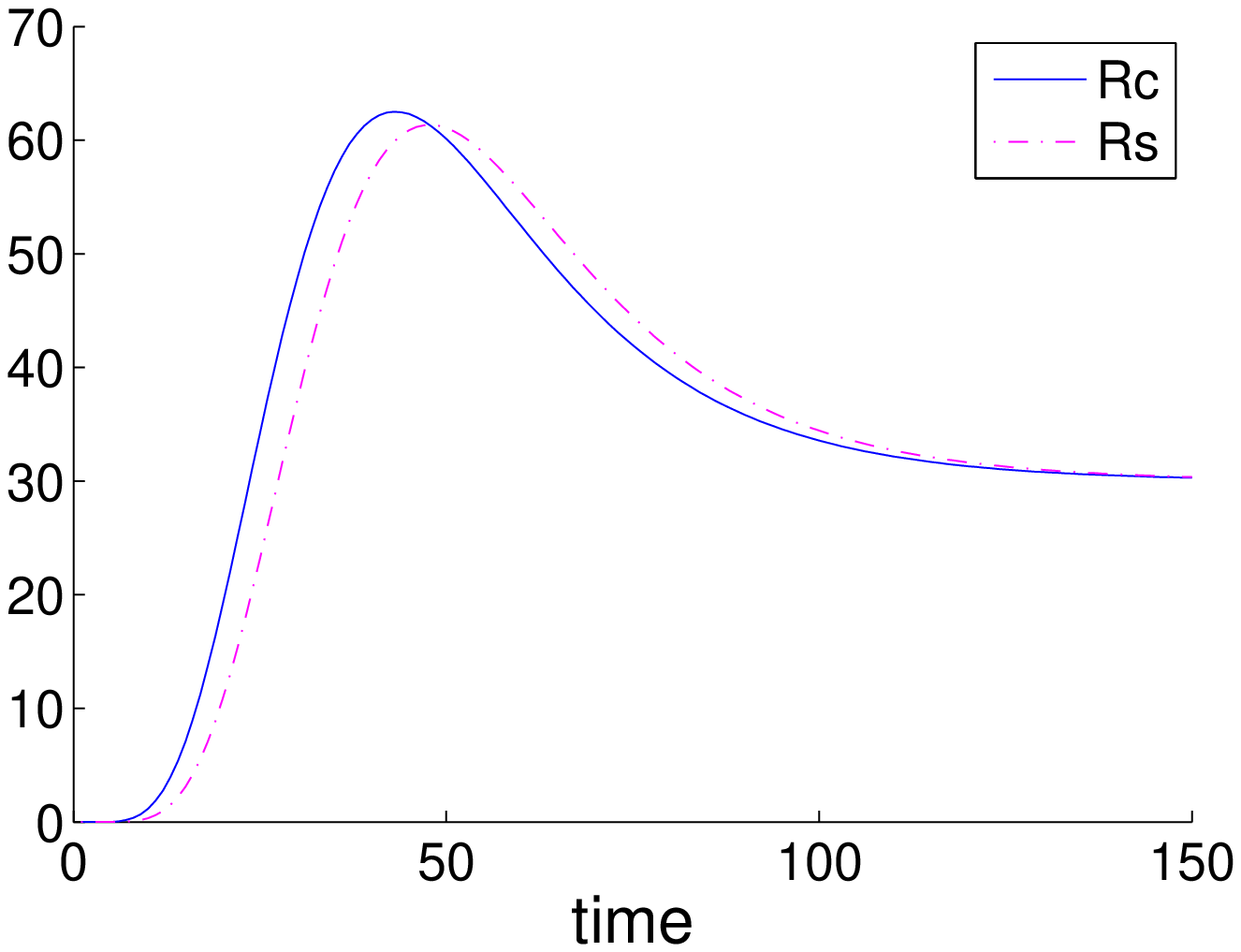}
\caption{Temporal filters $\Rc$ and $\Rs$.}
\label{fig:tempFilter}
\end{figure}

The input signal is a still-image which exists for a long time, hence $f(x,t) = f(x) \mathbbm{1}_{[0,\infty]}(t)$ where $f(x)$ is the still-image and $\mathbbm{1}$ is the indicator function such that $\mathbbm{1}_{[0,\infty]}(t) =1,$ if $0\leq t\leq \infty$, otherwise $0$. In this case, we obtain the following simplified convolution formula.

\begin{prop}
\label{pro:image}
 For a still-image $f(x,t) = f(x) \mathbbm{1}_{[0,\infty]}(t)$, the spatiotemporal convolution (\ref{eq:SpaTempConv}) turns into a spatial convolution:
\begin{equation}
\begin{array}{llll}
A(x,t) &=  \phi(x,t) \overset{x}{*} f(x),
\end{array}
\label{eq:SpaConv}
\end{equation}
where $\phi(x,t)$ is a spatial DoG filter weighted by two temporal filters $\Rc$ and $\Rs$:
\begin{equation}
\phi(x,t) = w_{c} \Rc  \Gc(x) -  w_{s} \Rs  \Gs(x),
\label{eq:nonSPAM}
\end{equation}
\begin{equation}
\Rc =  \intt  \Txt   dt',
\label{eq:tempCent}
\end{equation}
\begin{equation}
\Rs = \intt  \TExt dt'.
\label{eq:tempSur}
\end{equation}
\end{prop}

\begin{figure}[!ht]
\centering
\includegraphics[scale=0.39]{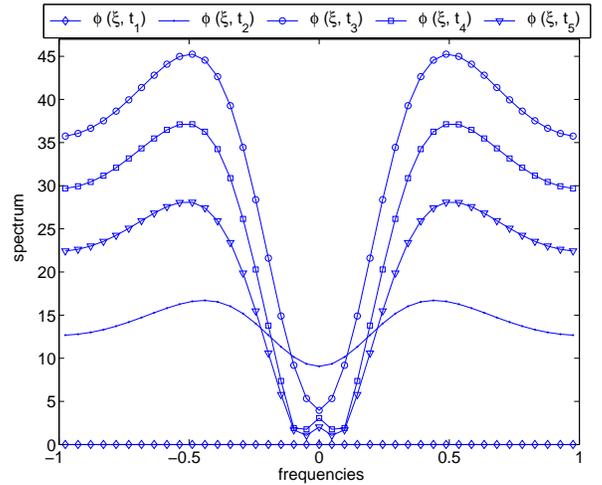}
\caption{The non-SPAM filter is a 2D spatially symmetric function. This figure shows a transversal cut of its spectrum for 5 different time samples of $\Rc$ and $\Rs$.}
\label{fig:SpatConvolution}
\end{figure}

\begin{figure}[!ht]
\begin{minipage}[b]{.30\linewidth}
  \centering
  \centerline{\includegraphics[width=3.3cm]{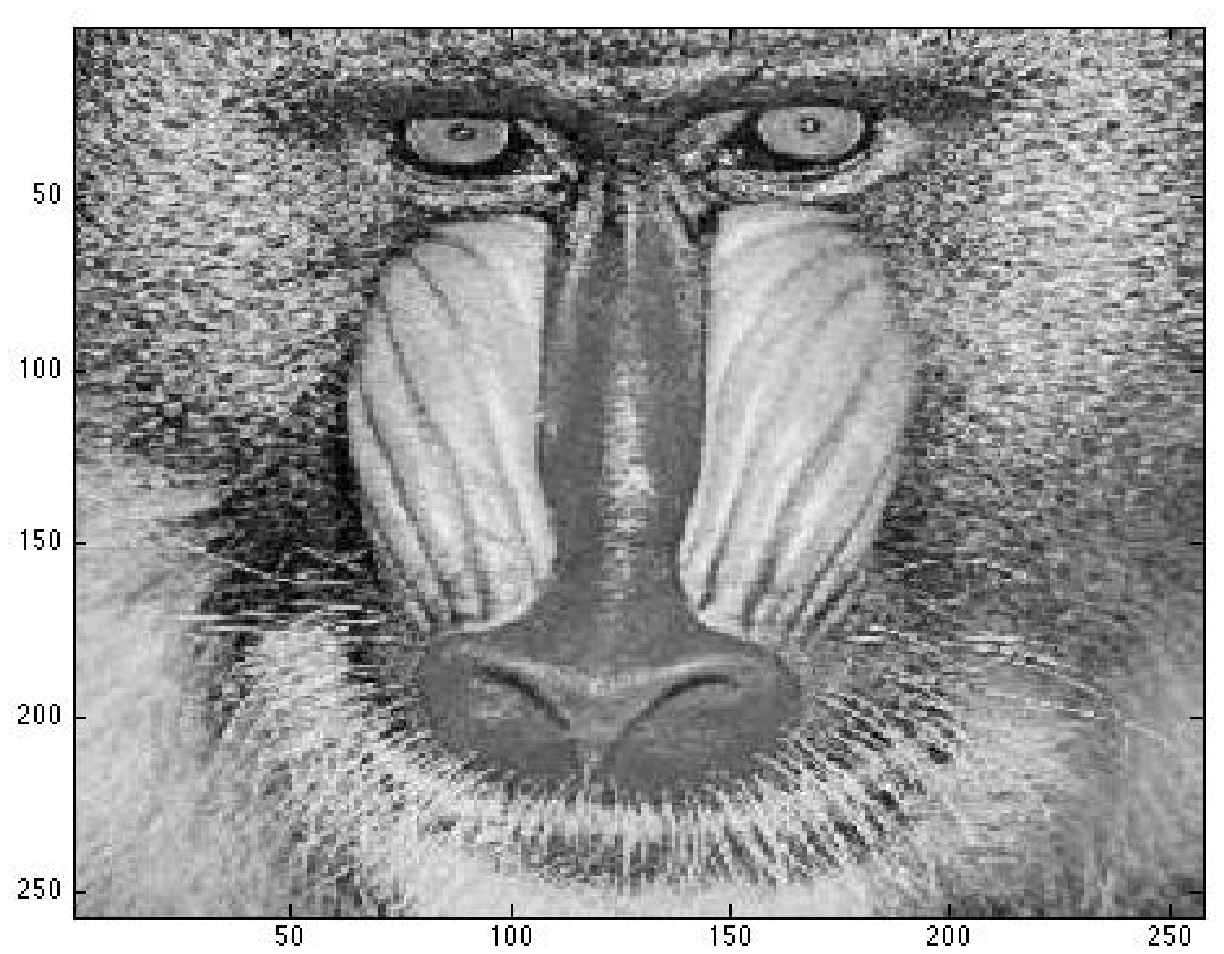}}
  \centerline{(a) Original Image}\medskip
\end{minipage}
\hfill
\begin{minipage}[b]{0.30\linewidth}
  \centering
  \centerline{\includegraphics[width=3.3cm]{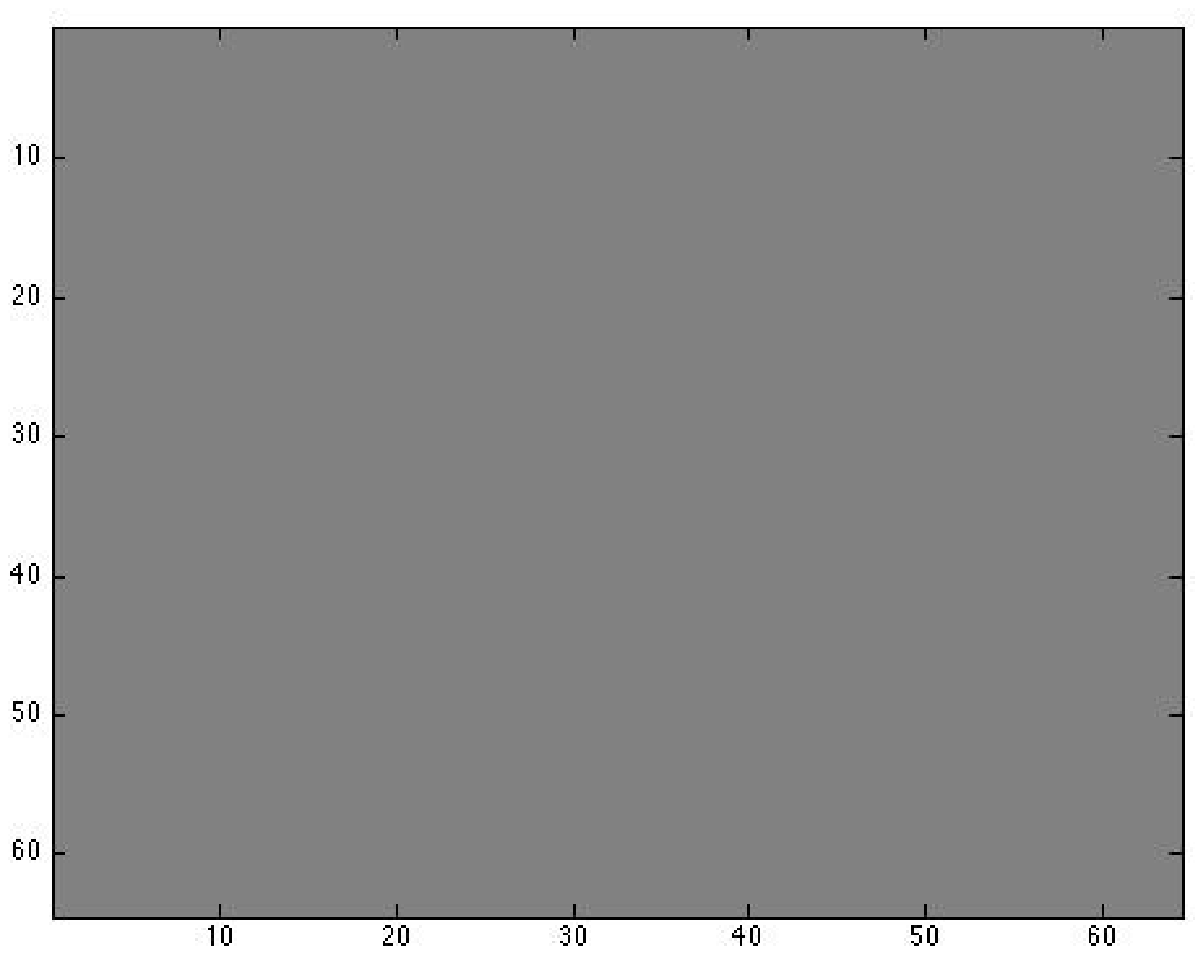}}
  \centerline{(b) Low-pass ($t_{1}$)}\medskip
\end{minipage}
\hfill
\begin{minipage}[b]{0.30\linewidth}
  \centering
  \centerline{\includegraphics[width=3.3cm]{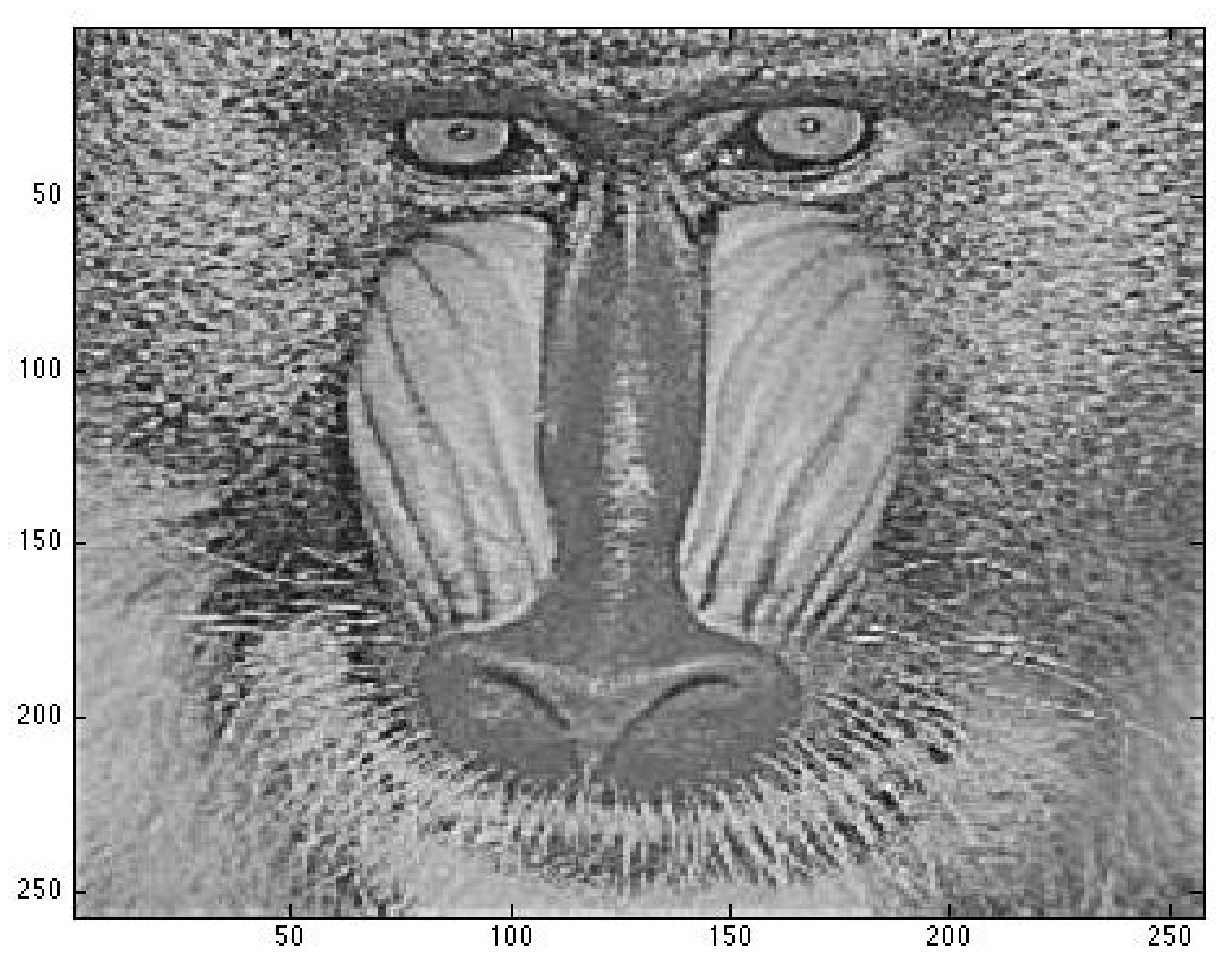}}
  \centerline{(c) Band-pass  ($t_{2}$) }\medskip
\end{minipage}
\begin{minipage}[b]{.30\linewidth}
  \centering
  \centerline{\includegraphics[width=3.3cm]{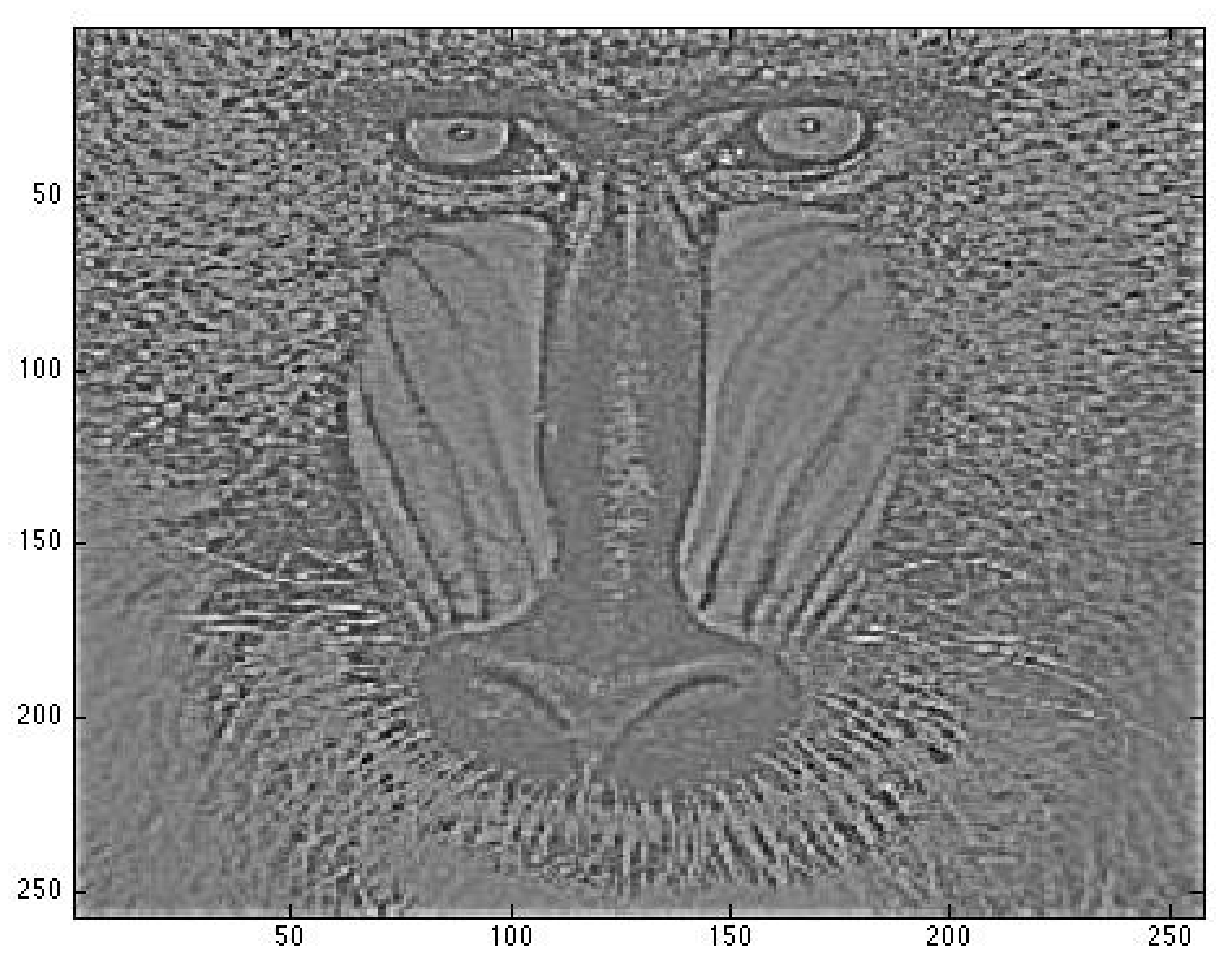}}
  \centerline{(d) Band-pass  ($t_{3}$)}\medskip
\end{minipage}
\hfill
\begin{minipage}[b]{0.30\linewidth}
  \centering
  \centerline{\includegraphics[width=3.3cm]{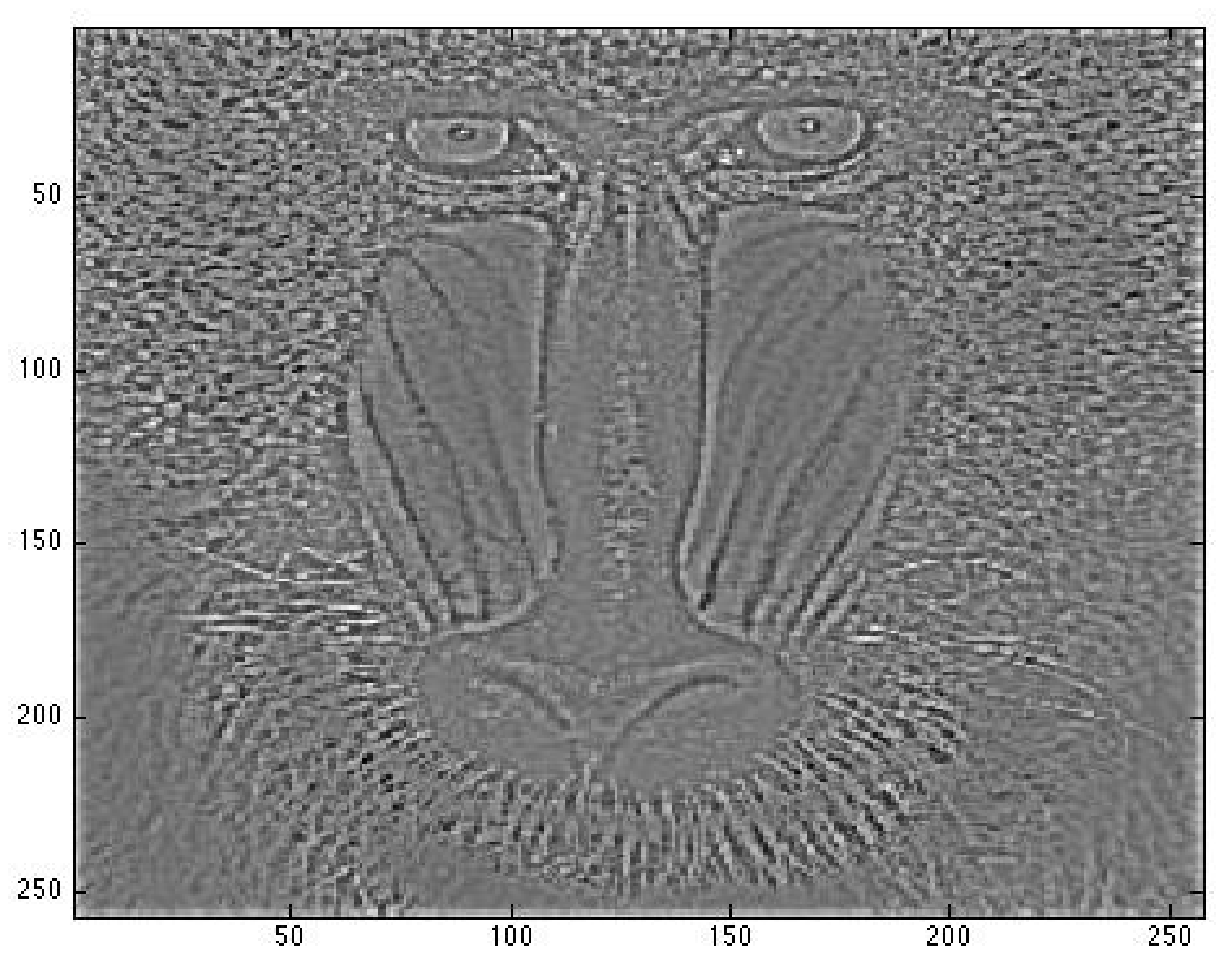}}
  \centerline{(e) Band-pass  ($t_{4}$)}\medskip
\end{minipage}
\hfill
\begin{minipage}[b]{0.30\linewidth}
  \centering
  \centerline{\includegraphics[width=3.3cm]{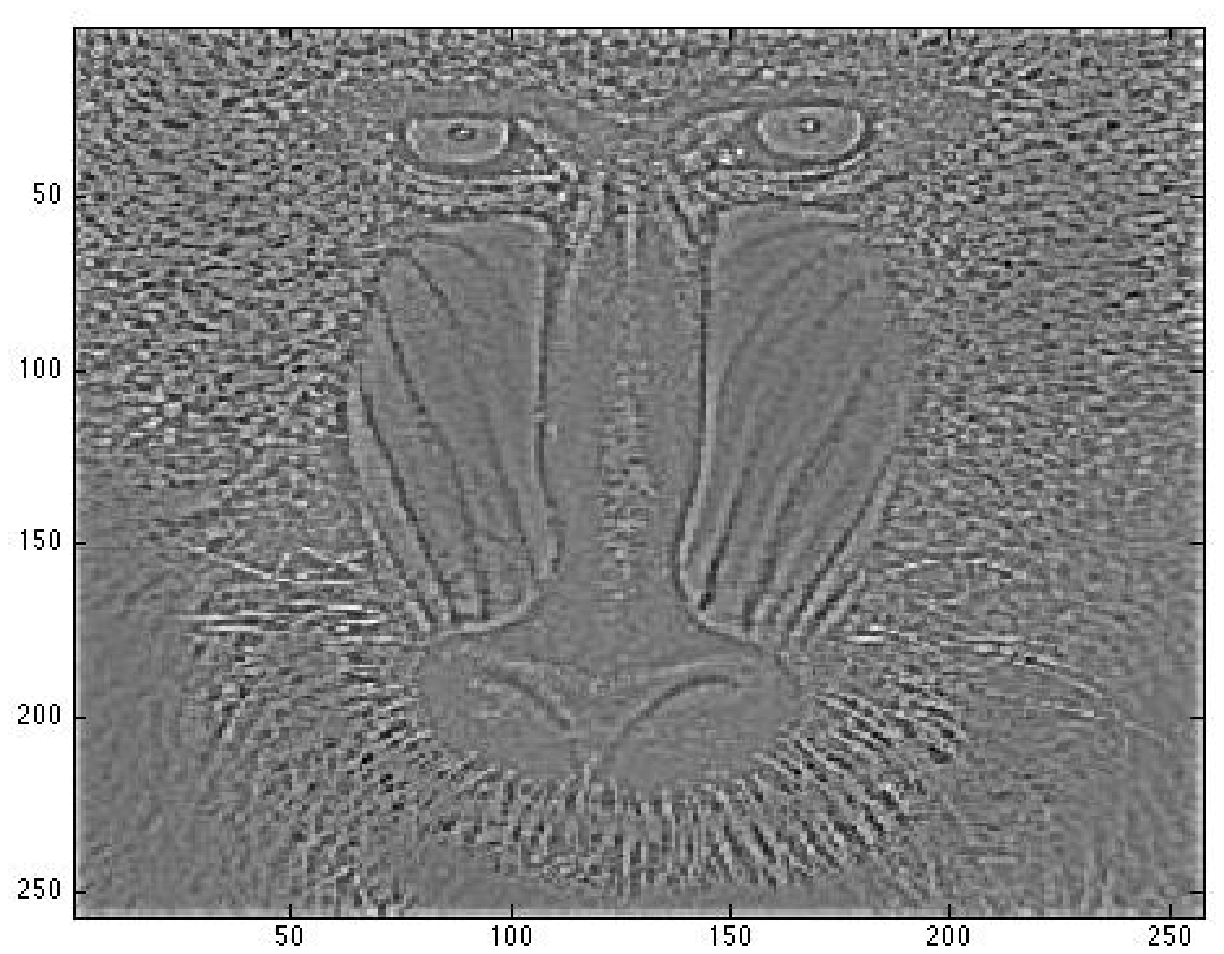}}
  \centerline{(f) Band-pass  ($t_{5}$)}\medskip
\end{minipage}
\caption{Image decomposition by the non-SPAM filter.}
\label{fig:SpatConvolutionResults}
\end{figure}

The above proposition is crucial for the reason that it enables the simplification and representation of the non-SPAM filter like a time-varying DoG. The DoG filters have been extensively studied in the past \cite{Field94,Cai97,Masmoudi12}. Proposition \ref{pro:image} shows that the retinal-inspired filter can be modeled by a spatial DoG filter which is multiplied by the temporal filters $\Rc$ and $\Rs$  (Fig. \ref{fig:tempFilter}), which act like weights and modify its spatial spectrum with respect to time (Fig. \ref{fig:SpatConvolution}). The non-SPAM filter is shown in Fig. \ref{fig:SpatConvolution} where the parameters have been tuned according to neuroscientic results which approximates the retinal spectrum and the speed of the retinal processing: $\tau_{C} = 20.10^{-3}, \tau_{s} = 4.10^{-3}, \tau_{G} = 5.10^{-3}, w_{S} = 1, w_{C} = 0.75, \sigma_{c} = 0.5, \sigma_{s} = 1.5$. One can notice that, after a while, both the temporal filters $\Rs$ and $\Rc$ converge to the same value, which is established in the following proposition.
\begin{prop}
The filter $\phi(x,t)$ is a continuous and infinitely differential function such that $\phi(x,0) = 0$ and 
$$
\displaystyle \lim_{t\rightarrow +\infty} \phi(x,t) = \phi(x)
$$ 
where $\phi(x)$ is a DoG filter.
\end{prop}
In practice, $\phi(x,t)$ almost converges within a short time delay $\Delta t$. Hence, we assume that $\phi(x,\Dt) \approx \phi(x)$ for all $x$. Hence, the non-SPAM filter only evolves during the time interval $[0,\Dt]$. The time when this convergence almost occurs marks the end of the spatial evolution of the non-SPAM filter. In other words, the non-SPAM filter is capable to decompose the input image into different spatial subbands and extract information for a certain period of time $\Dt$. After $\Dt$ all the necessary information has already been selected (Fig. \ref{fig:SpatConvolutionResults}) .

\section{non-SPAM Frame}
\label{sec:frame}

The goal of this section is to prove that the non-SPAM filter is invertible such that we are able to reconstruct the input image. For this reason, we establish that the non-SPAM filter has a frame structure \cite{Masmoudi12,Kovacevic08}. For numerical purpose, we need to discretize the non-SPAM filter. Let $x_{1},\ldots,x_{n} \in \Rn$ and $t_{1},\ldots,t_{m}\in \R^{+}$ be some sets of spatial and temporal sampling points. As a consequence, the continuous spatial convolution is approximated by the discrete convolution:
\begin{eqnarray}
A(x_{k},t_{j}) &=& \phi(x_{k},t_{j}) \circledast f(x_{k})\nonumber  \\
&=& \SUM{i=1}{n} \DKxt \Dfx,\label{discSpatConv}
\end{eqnarray}
for all $k$ and $j$. Let $\varphi_{k,j}$ be the row vector of $\R^n$ defined by
$$
\varphi_{k,j}=\Big(\phi(x_k-x_{1},t_j),\ldots,\phi(x_k-x_{n},t_j)\Big)
$$
and $\Phi$ be the family of all these vectors:
\begin{equation}
\Phi = \Big\lbrace \varphi_{k,j} \Big\rbrace_{1\leq k \leq n,1\leq j\leq m}.
\label{frame}
\end{equation}
Let $f=\left(f(x_1),\ldots,f(x_n)\right)$ be the discretized image and  $\|f\|$ its Euclidean norm.
Let us denote $\tilde{\phi}_{t_j}(\xi)$ the discrete Fourier transform of the vector $\left(\phi(x_1,t_j),\ldots,\phi(x_n,t_j)\right)$.
\begin{prop}
The family of vectors $\Phi$  is a frame i.e. there exist two scalars $0 < \alpha \leq \beta < \infty$ such that:
\begin{equation}
\begin{array}{l l l l l l l }
\alpha \|f \|^{2} &\leq  \SUM{j=1}{m} \SUM{k=1}{n} | A(x_{k}, t_{j}) | ^{2} &\leq \beta \|f\|^{2},
\end{array}
\label{eq:FrameMallat}
\end{equation}
where 
$$
\alpha = \min_{\xi}\Big\lbrace \displaystyle \frac{1}{n} \SUM{j=1}{m} \Big|  \tilde{\phi}_{t_{j}}(\xi) \Big|^{2}  \Big\rbrace >0,
$$ 
$$ 
\beta =  \SUM{j=1}{m} \SUM{k=1}{n} \SUM{i=1}{n}  \phi^{2}(x_{k} - x_{i},t_j).
$$
\label{prop:frame}
\end{prop}
The proof of Proposition \ref{prop:frame} is omitted due to the lack of place.

\section{Progressive Reconstruction}
\label{sec:reconstruction}

The progressive reconstruction consists in computing an estimate $\hat{f}_{t_m}$ of the discretized image $f$ at time $t_{m}$ by using a limited amount of coefficients. In fact, at time $t_{m}$, all the coefficients of the non-SPAM frame has been computed but the reconstruction only exploits a small amount of them. We use the term progressive because the quality of the reconstruction increases as the amount of the coefficients in use increases. 

When the decoder knows the total number of coefficients, he can reconstruct perfectly the input signal. Let us define
$\tilde{A} = [\tilde{A}_{t_{1}}, \ldots, \tilde{A}_{t_{m}}]$ as a vector of size $nm$ and  $\Phi = [\phi_{1}, \ldots, \phi_{m}]$ a matrix of size $nm\times n$, where
\begin{equation}
\tilde{A}_{t_{j}} = \displaystyle \left[
\begin{array}{c c}
\tilde{A}(x_{1},t_{j}) \\
\vdots \\
\tilde{A}(x_{n},t_{j})
\end{array}
\displaystyle \right]\;\;\;\mbox{and}\;\;\;
 \phi_{j}= \displaystyle \left[
 \begin{array}{c}
 \varphi_{1,j}  \\
 \vdots \\
 \varphi_{n,j}
\end{array}
\displaystyle \right].
\end{equation}
At time $t_m$, the estimate $\hat{f}_{t_m}$ is given by
\begin{equation}
\tilde{f}_{t_{m}} =  {( \Phi^{\top} \Phi )}^{-1} \Phi^{\top} \tilde{A},
\label{eq:Recon}
\end{equation}
where $M^{-1}$ denotes the inverse of a matrix $M$ and $M^\top$ denotes its transpose. The dual frame, which is necessary to have a perfect decoding at time $t_m$ \cite{Masmoudi12,Kovacevic08}, is $( \Phi^{\top} \Phi )^{-1} \Phi^{\top}$. Instead of computing the above matrix operator which can be time consuming and resource demanding, we can note that (\ref{eq:Recon}) is a solution of the following least squares problem:
\begin{equation}
\tilde{f}_{t_{m}}=\arg\min_{f\in\R^n}\left( \sum_{j=1}^{m}\| \phi_{j} \circledast f - \tilde{A}_{t_{j}} \|^2 \right).
\label{eq:least-squares}
\end{equation}
This minimization problem can be easily solved by using a gradient descent algorithm.

The progressive reconstruction of the decoder is computed by selecting for each time bin the same percentages of spatial coefficients which are used to produce the reconstructed image. The decoder sorts in a descending order the total amount of coefficients for each time bin, and then it extracts the coefficients with the highest energy omitting the rest of the data. Obviously, the numerical results show that the reconstruction is better as the number of extracted coefficients is larger. This approach is based on the Rank-Order-Coding (ROC) model which is proposed in \cite{Thorpe01}.

\section{Results}
\label{sec:results}
This section gives the numerical results of the progressive decoding introduced in the previous section. 
The parameters of the simulation are given in Fig. \ref{fig:SpatConvolution}. The image is composed of $64\times 64$ pixels. 
%
We define MSE$(f,\hat{f}_{t_m})=||f-\hat{f}_{t_m}||^2/n$ the mean square error which measures the distortion between the original image $f$ and the reconstructed image $\hat{f}_{t_{m}}$. 

\begin{figure}[!ht]
\begin{minipage}[b]{.40\linewidth}
  \centering
  \centerline{\includegraphics[width=4.3cm]{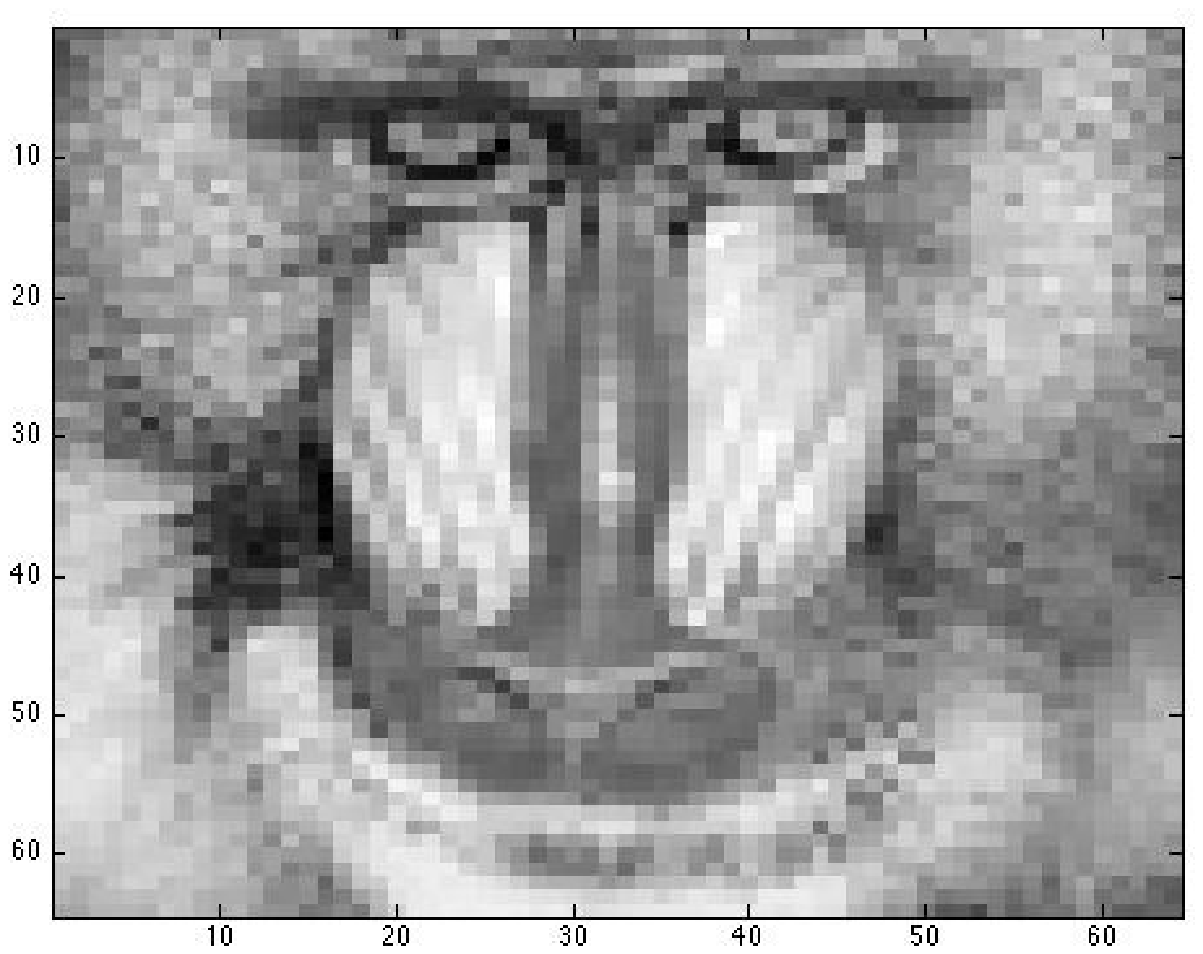}}
  \centerline{\begin{tabular}{c c}
  (a) Original \\ Image (64x64)
  \end{tabular}}  \medskip
\end{minipage}
\hfill
\begin{minipage}[b]{0.40\linewidth}
  \centering
  \centerline{\includegraphics[width=4.3cm]{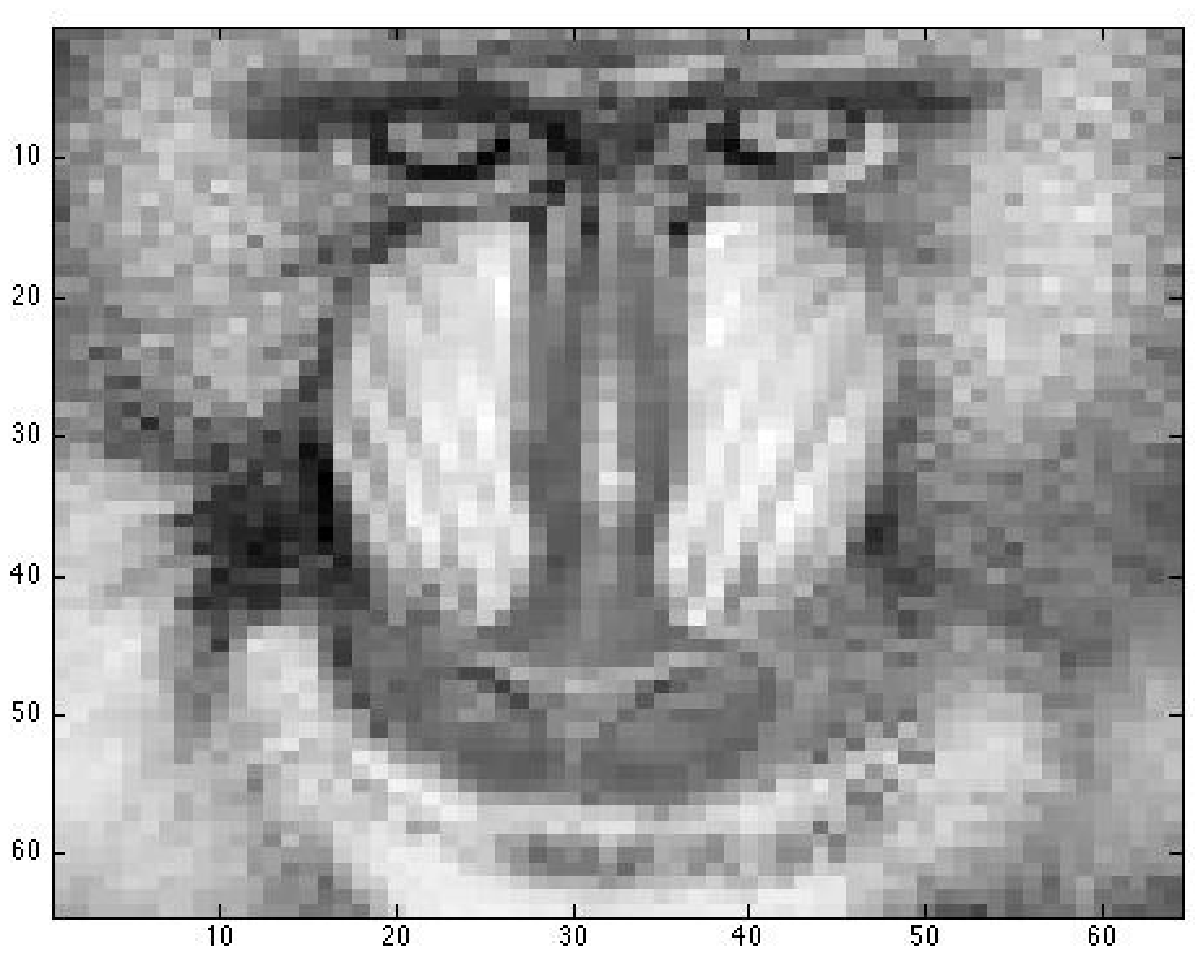}}
  \centerline{ \begin{tabular}{c c}
  (b) MSE$(f,\hat{f}_{t_{m}})$= \\ 0.3884 \end{tabular} }\medskip
\end{minipage}
\caption{Frame-based reconstruction.}
\label{fig:Results}
\end{figure}

\begin{figure}[!ht]
\begin{minipage}[b]{.30\linewidth}
  \centering
  \centerline{\includegraphics[width=3.3cm]{64x64.eps}}
  \centerline{\begin{tabular}{c c}
  (a) Original \\ Image  \\ (64x64)
  \end{tabular}}  \medskip
\end{minipage}
\hfill
\begin{minipage}[b]{0.30\linewidth}
  \centering
  \centerline{\includegraphics[width=3.3cm]{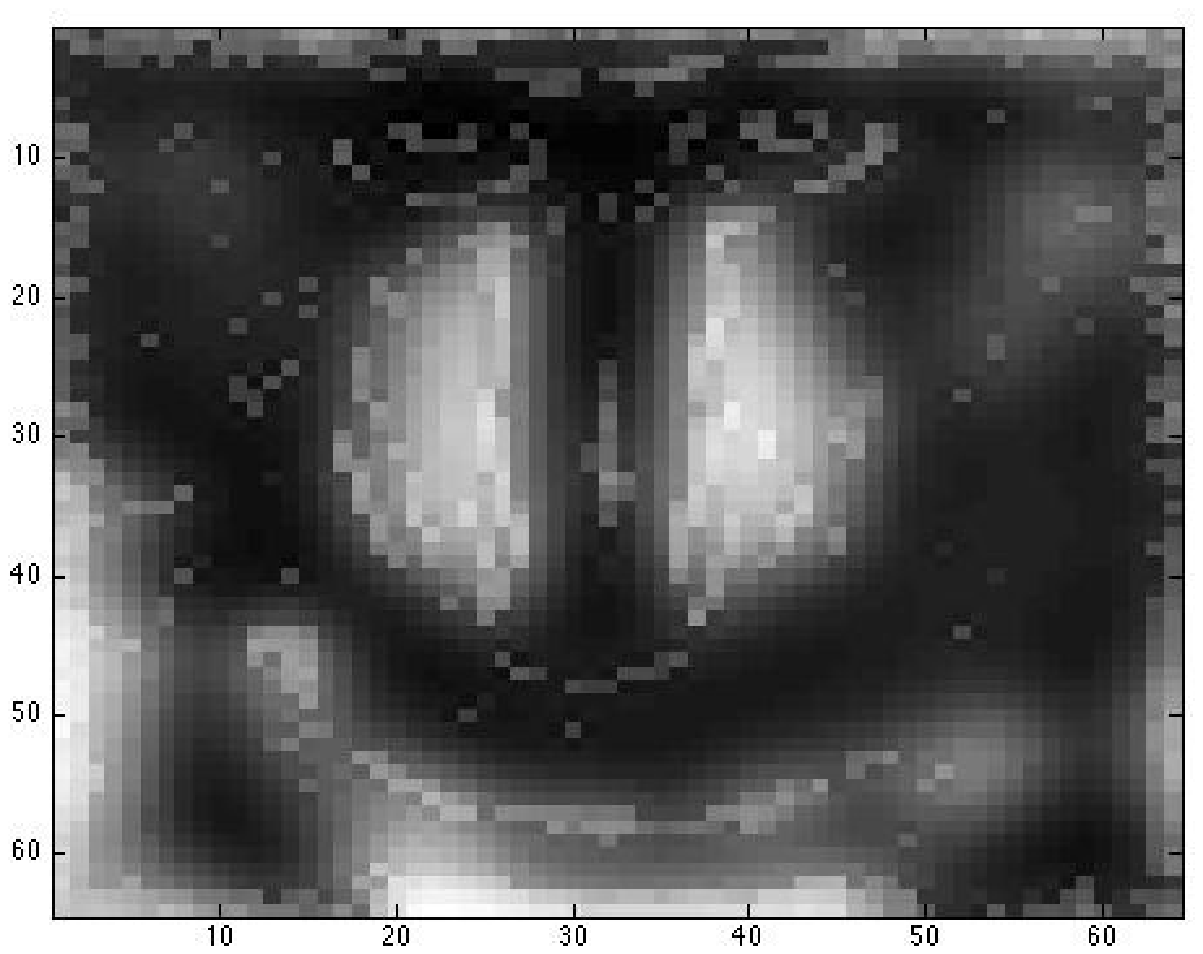}}
  \centerline{\begin{tabular}{c c}
  (b) (20\%) \\ MSE$(f,\hat{f}_{t_{m}})$ = \\ 1.2120e+04 
  \end{tabular}}  \medskip
\end{minipage}
\hfill
\begin{minipage}[b]{0.30\linewidth}
  \centering
  \centerline{\includegraphics[width=3.3cm]{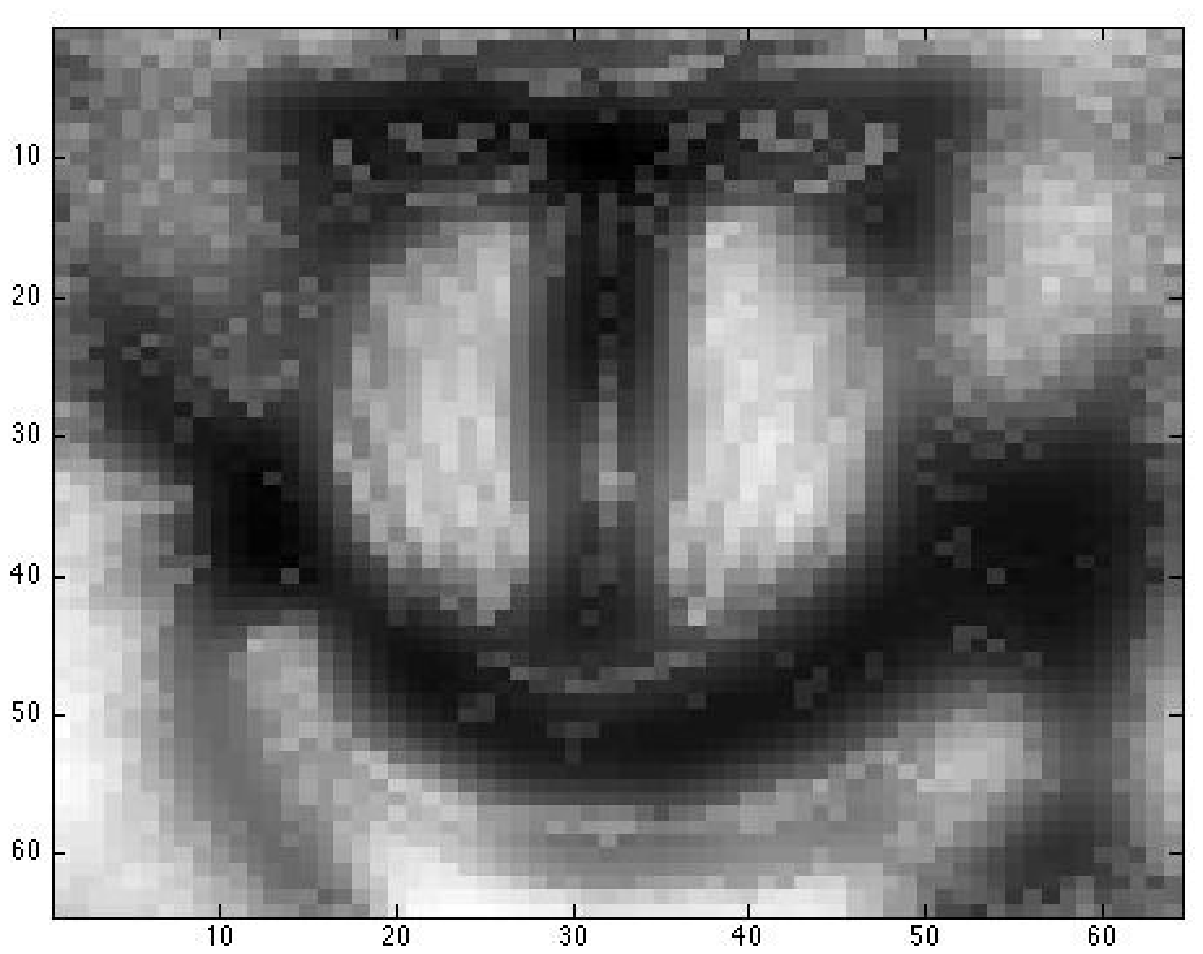}}
  \centerline{ \begin{tabular}{c c}
  (c) (40\%) \\  MSE$(f,\hat{f}_{t_{m}})$= \\6.2147e+03 
   \end{tabular} }\medskip
\end{minipage}
\begin{minipage}[b]{.30\linewidth}
  \centering
  \centerline{\includegraphics[width=3.3cm]{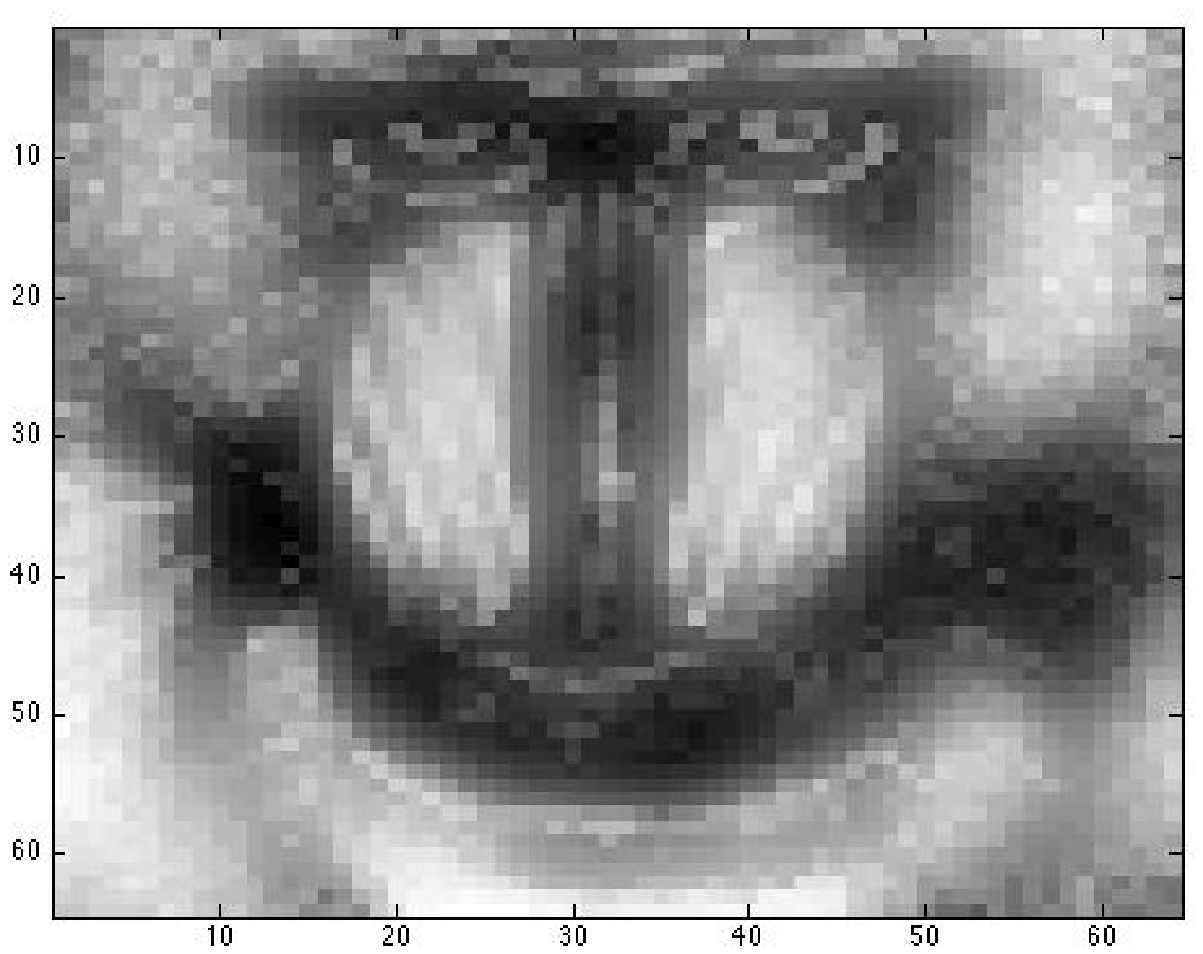}}
  \centerline{ \begin{tabular}{c c}
  (d) (60\%) \\ MSE$(f,\hat{f}_{t_{m}})$= \\ 2.5820e+03 
  \end{tabular} }\medskip
\end{minipage}
\hfill
\begin{minipage}[b]{0.30\linewidth}
  \centering
  \centerline{\includegraphics[width=3.3cm]{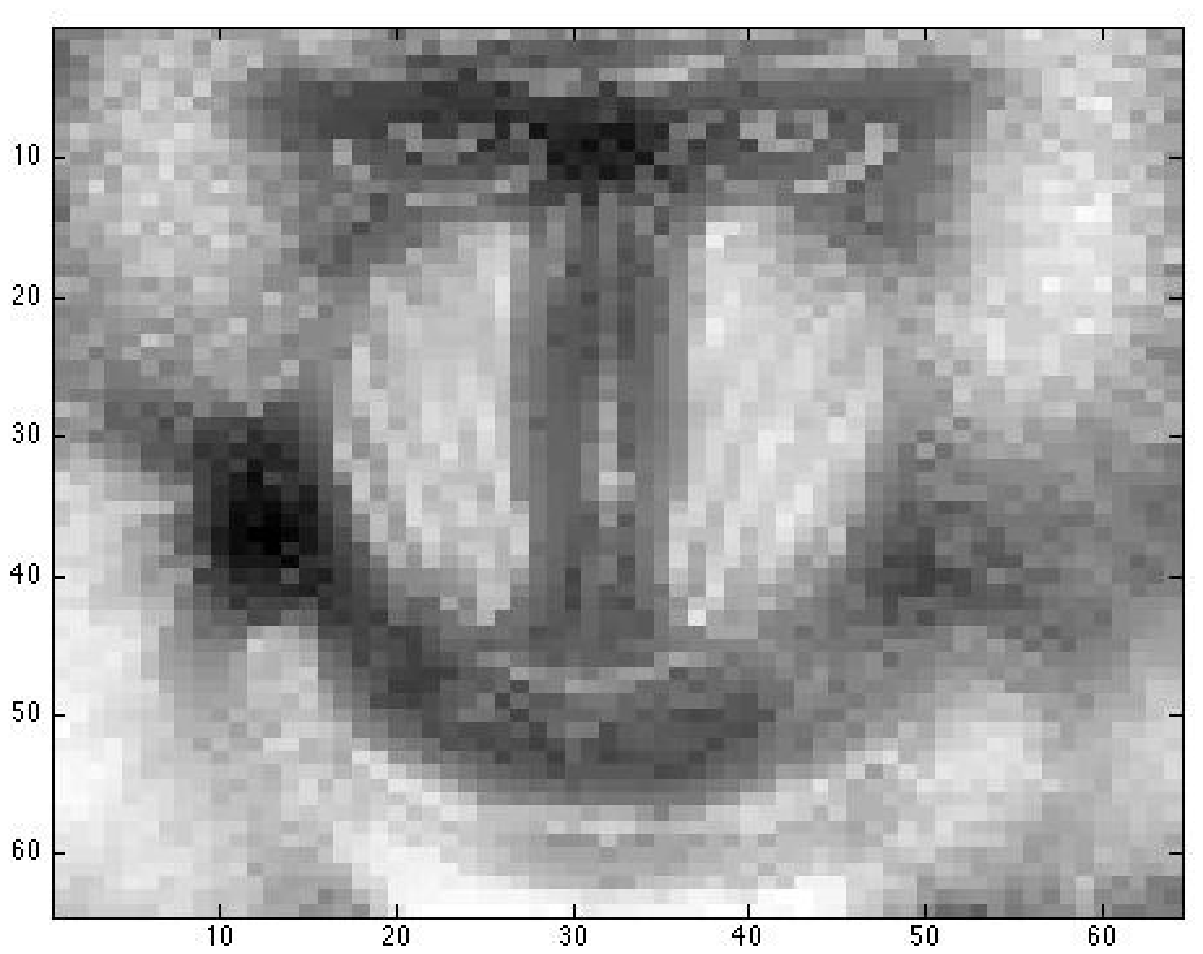}}
  \centerline{ \begin{tabular}{c c}
  (e) (80\%) \\ MSE$(f,\hat{f}_{t_{m}})$= \\ 664.4727 
  \end{tabular} }\medskip
\end{minipage}
\hfill
\begin{minipage}[b]{0.30\linewidth}
  \centering
  \centerline{\includegraphics[width=3.3cm]{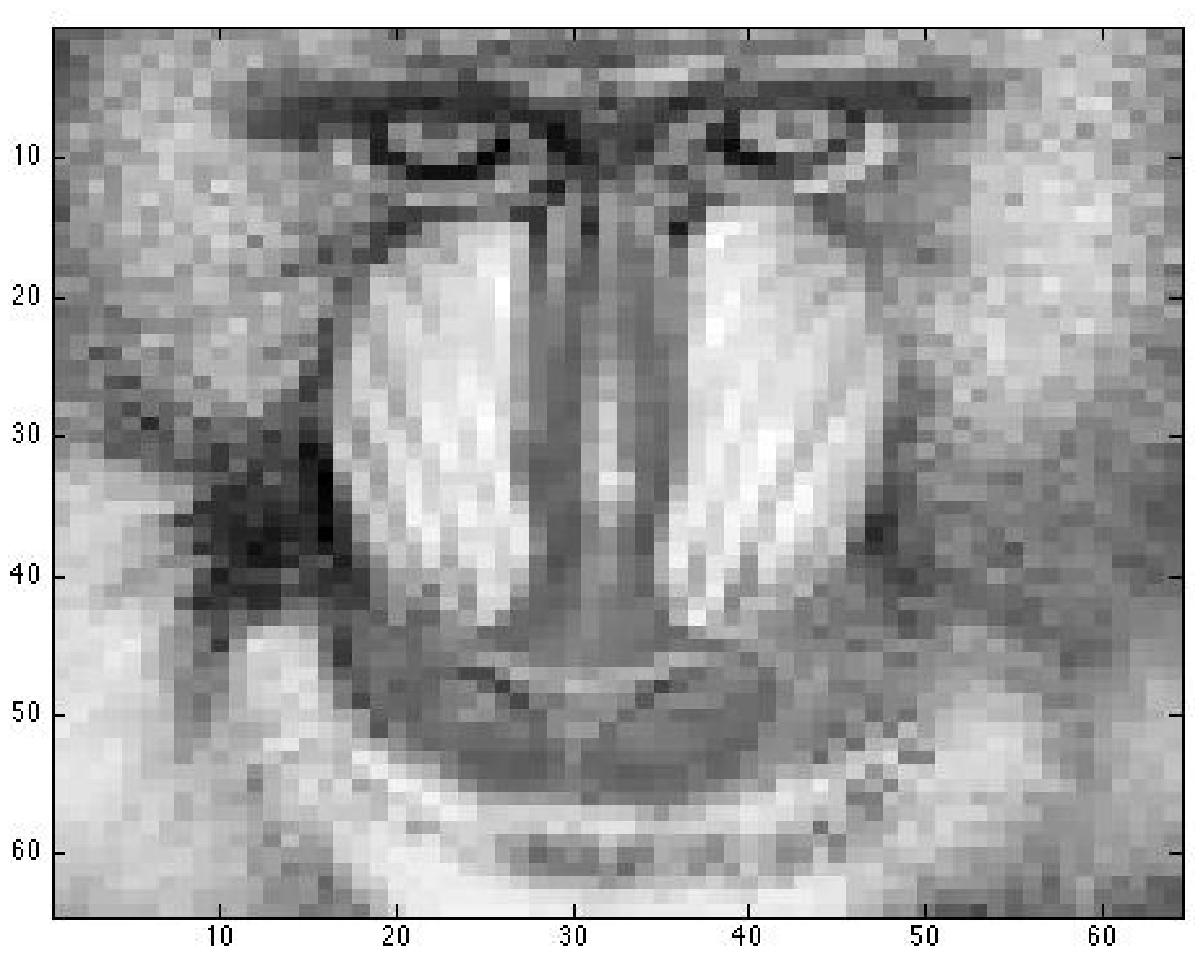}}
  \centerline{ \begin{tabular}{c c}
  (f) (100\%) \\ MSE$(f,\hat{f}_{t_{m}})$= \\ 0.3884 
   \end{tabular} }\medskip
\end{minipage}
\caption{Progressive reconstruction based on ROC model.}
\label{fig:LIFResults}
\end{figure}

Fig. \ref{fig:Results} shows the optimal reconstruction when all the coefficients are used. 
The progressive reconstruction which is based on the ROC model is given in Fig. \ref{fig:LIFResults}. It is obvious that while the number/percentage of coefficients increases, the quality of the reconstruction improves. However, even for a very small percentage of the total amount of coefficients (i.e., 40\%), the sharpness of the reconstructed image allows to recognize the basic structures of the input image.

\section{Conclusion}
\label{sec:conclusion}

In this document, we have introduced a novel non-separable spatiotemporal filter (non-SPAM) based on a retinal model. This filter has a time-varying behavior. The progressive reconstruction exploits the rank-order-coding model to reconstruct the image by using a limited number of coefficients. For further study, we aim to extend this filtering result on a video stream. 

\bibliographystyle{IEEEbib}
\bibliography{refs}

\end{document}